\documentclass[aps,prl,twocolumn,groupedaddress,showpacs]{revtex4}
\usepackage{graphicx,amsmath,amssymb,amsfonts,latexsym,color,dcolumn,bm}
\usepackage{verbatim}

\newcommand{\curlybraket}[1]{ \left\{ {#1} \right\} }
\newcommand{\squarebraket}[1]{ \left[ {#1} \right] }
\newcommand{\roundbraket}[1]{ \left( {#1} \right) }

\newcommand{\ket}[1]{\left|{#1}\right\rangle}

\newcommand{\req}[1]{Eq.~(\ref{#1})}

\newcommand{\beq}{\begin{equation}} 
\newcommand{\eeq}{\end{equation}}
\newcommand{\bqa}{\begin{eqnarray}} 
\newcommand{\eqa}{\end{eqnarray}}

\newcommand{\dg}{^{\dagger}}

\newcommand{\avg}[1]{\left\langle{#1}\right\rangle}

\def\*#1{\mathbf{#1}}

\begin{document}

\title{Ultrafast QND measurements based on diamond-shape artificial atom}

\author{I. Diniz, E. Dumur, O. Buisson and A. Auff\`eves}

\affiliation{Institut N\'eel, C.N.R.S.- Universit\'e Joseph Fourier, BP 166, 38042
Grenoble-cedex 9, France}

\date{\today}

\begin{abstract}

We propose a Quantum Non Demolition (QND) read-out scheme for a superconducting
artificial atom coupled to a resonator in a circuit QED architecture, for which we
estimate a very high measurement fidelity without Purcell effect limitations. The device
consists of two transmons coupled by a large inductance, giving rise to a diamond-shape
artificial atom with a logical qubit and an ancilla qubit interacting through a
cross-Kerr like term. The ancilla is strongly coupled to a transmission line resonator.
Depending on the qubit state, the ancilla is resonantly or dispersively coupled to the
resonator, leading to a large contrast in the transmitted microwave signal amplitude.
This original method can be implemented with state of the art Josephson parametric
amplifier, leading to QND measurements in a few tens of nanoseconds with fidelity as
large as 99.9\%.

\end{abstract}


\maketitle

Superconducting circuits have demonstrated in the last decade their high ability to
perform coherent quantum experiments\cite{Korotkov_QIP2009,Clarke_Nature2008}.
Relaxation $T_1$ and coherence $T_2$ times are continuously increasing
\cite{Paik_PRL2011}. In addition, these quantum systems benefit from very strong
coupling with the electromagnetic field, and potential scalability. Finally the circuit
parameters that define the quantum dynamics are tunable and adjustable on demand, which
makes them very promising candidates to process quantum information on chip. In this
framework, the ability to perform ultrafast single shot read-out of a quantum bit is
highly desirable. Up to now, high one shot fidelity was obtained by switching quantum
measurements using escape process \cite{Lucero_PRL2008}, the intrinsic drawback of this
method being its destructiveness. Quantum Non Demolition (QND) measurements are
performed by coupling the qubit dispersively to a resonator \cite{Wallraff_PRL2005}. The
qubit acts as a state-dependent refractive index that shifts the cavity frequency, and
the measurement is performed by probing the resonator with an external microwave. QND
character is preserved as long as one remains in the dispersive regime, keeping the
photon population $\bar{n}$ of the resonator below a critical
value\cite{Boissonneault_PRA2009}, and also limiting the incident power. Low temperature
amplifiers have thus to be used to reach high fidelity. On the other hand, using
non-linear resonator and bifurcation\cite{Siddiqi_PRB2006,Mallet_naturephys2009} or
Jaynes-Cummings non linearity\cite{Reed_PRL2010}  allows to reach one shot high fidelity
read-out, but at the price of lower QND fidelity. Thanks to recent advances in
parametric amplification using Josephson junction circuits
\cite{Clerk_RMP2010,Bergeral_Nature2010,Roch_PRL2012}, single shot read-out has been
demonstrated, allowing to observe quantum jumps in superconducting artificial
atoms\cite{Vijay_PRL2011}, high fidelity
read-out\cite{Johnson_PRL2012,Riste_PRL2012,Hatridge_Science2013} and ever-persisting
Rabi oscillations\cite{Vijay_Nature2012}. However this measurement scheme still requires
several hundred nanoseconds measurement time to reach high fidelity. Consequently,
further improvements are necessary in order to reach very high fidelity measurements in
a few ten's of nanoseconds. New quantum measurement protocols inspired by ion traps and
quantum optics were recently proposed with this
purpose\cite{Solano_PRB2010,Kumar_PRB2010}.

Here we propose an original method to realize ultrafast QND measurements of a qubit with
large resonator linewidth and measurement bandwidth, while preserving high fidelity. Our
system is a resonator coupled to a four level diamond-shape atom, which can be seen as
two qubits coupled by crossed Kerr interaction. In this picture, the first qubit is the
one we read out, while the second qubit plays the role of an ancilla whose frequency
depends on the first qubit state.  The resonator is not coupled to the qubit but only to
the ancilla. This huge difference with respect to previous experiments induces important
consequences on the qubit and on the resonator properties. First, Purcell effect between
the qubit and the cavity is absent, and the readout performance is independent from the
detuning between the qubit and the resonator. Second, the present proposal allows to
eliminate the harsh constraint on the amplification and resonator bandwidth, allowing to
reach fast, one-shot, high fidelity QND read-out of our qubit even with the present day
amplifier technology.

\begin{figure}[htb]
\begin{center}
\includegraphics[width=8cm]{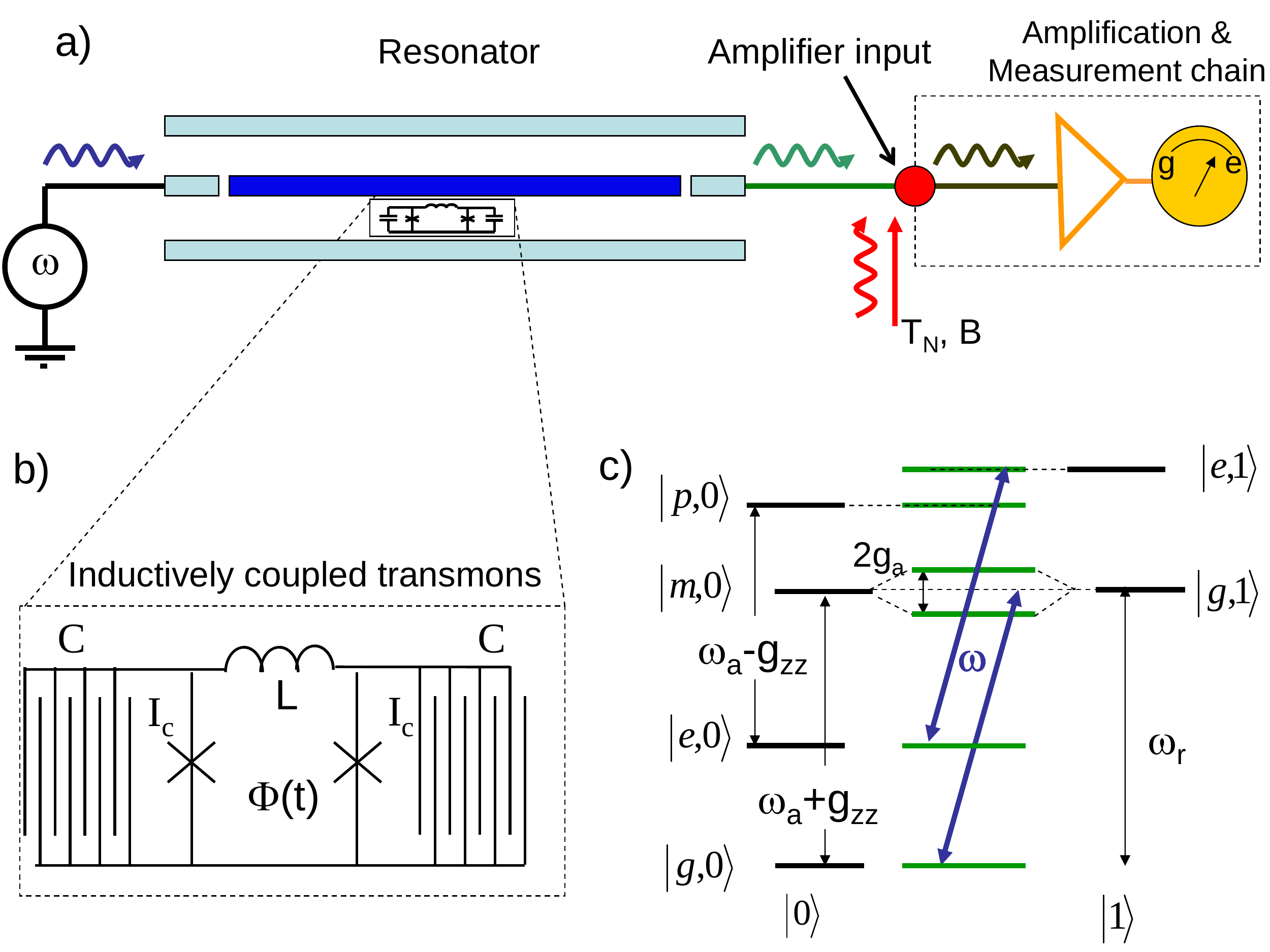}
\caption{(a) Schematic circuit for a fast QND read-out. The incident microwave signal of frequency $\omega$
is transmitted through the resonator coupled to the
superconducting artificial atom. The transmitted signal is
amplified and homodyne detected. The noise
amplifier is illustrated by additional microwave source at the amplifier input. (b) The artificial
atom, realized by two transmons coupled by a large inductance. (c)Energy spectrum of the uncoupled (left and right) and dressed (center) diamond shape
artificial atom-resonator states when $\omega_r=\omega_a + g_{zz}$. The blue arrows
indicate the injected microwave frequency when $\omega=\omega_r$.} \label{fig:1}
\end{center}
\end{figure}

We exemplify hereafter this method with a diamond-shape artificial atom consisting of
two transmons coupled by a large inductance. The system under study here is pictured in
Fig.\ref{fig:1}, in which the resonator is schematized by a coplanar wave guide
resonator, but our result can be applied to 3D cavities or lumped element resonators.
The Josephson energy $E_J=\Phi_0 I_c/2\pi$ and the charging energy $E_c=e^2/2C$, with
$I_c$ and $C$ the critical current and the capacitance per junction, are fixed to verify
the typical ratio $E_J/E_c\approx 50$ of the transmons limiting the decoherence effects
\cite{Koch_PRA2007}. Such devices can be described by an anharmonic oscillator with two
degrees of freedom. Its quantum description is the same than the one performed in
Ref.\cite{Lecocq_PRL2011} for a dc SQUID in the particular case of zero-current bias or
in Ref.\cite{Hu_PRB2011,Neumeier_arXiv2012} for two transmons coupled by a SQUID. It
gives rise to two orthogonal modes, the symmetric and antisymmetric
mode\cite{Lecocq_PRL2011}. When the anharmonicity is strong enough, we can consider just
the first two levels of the two modes. The system is then reduced to two coupled
two-level systems. The two first quantum states of the symmetric mode, $\ket{g}$ and
$\ket{e}$, provide the logical qubit $\sigma_z^{qb}$. The second two level system
corresponds to the two first quantum states of the antisymmetric mode, and it will be
used as an ancilla $\sigma_z^{a}$ for the quantum measurement. In the absence of
coupling between the two modes, the transition frequencies of the qubit and ancilla are
given by of $\omega_{qb}$ and $\omega_a$, respectively. Hereafter we restrict our study
to the working point given by zero-flux bias which constitutes an optimal point for the
artificial atom. The coupling between the two systems is reduced to a longitudinal
interaction term $\sigma_z^{qb}\sigma_z^{a}$ with a strength given by $\hbar g_{zz}=
E_c/(\sqrt{1+2E_L/E_J})$\cite{Lecocq_PRL2012} where $E_L=(\Phi_0/2\pi)^2/L$ with $L$ the
SQUID inductance. This term can be viewed as an analogue of a cross-Kerr term between
the two quantum systems, leading to a conditional energy transition of the ancilla which
depends on the quantum state of the qubit $\ket{g}$ and $\ket{e}$. The respective
frequencies of the transitions $\ket{g}\rightarrow\ket{m}$ and
$\ket{e}\rightarrow\ket{p}$ are $\hbar (\omega_a + g_{zz}) $ and $\hbar (\omega_a -
g_{zz}) $. The artificial atom inside a coplanar resonator is described by the following
Hamiltonian, written in the rotating wave approximation:
\begin{equation}\begin{split}\label{eq:Hamiltonian0}
H_{free} = \hbar \omega_{qb} \sigma_z^{qb}/2 + \hbar (\omega_a - g_{zz}\sigma_z^{qb}) \sigma_z^{a}/2 \\
 +\hbar \omega_{r} (a\dg a+1/2) - i \hbar g_a (a \sigma_{+}^{a} - a\dg \sigma_{-}^{a})
\end{split}
\end{equation}
The first three terms describe the artificial atom and the
fourth describes the resonator of frequency $\omega_r$. In the following we choose the frequency condition between the
resonator and the ancilla: $\omega_r=\omega_a + g_{zz}$. The last term couples the
resonator and the ancilla when the artificial atom is localized at the center of the
resonator. Indeed, at this particular place, the quantum fluctuation of the flux is
maximal and the voltage fluctuations are reduced to zero. Because of zero-flux bias working point, the qubit is not affected by flux
fluctuations, leading to zero-coupling between the resonator and the qubit. This way,
$\sigma_z^{qb}$ commutes with the Hamiltonian of the system, ensuring the
non-destructive character of the measurement whatever the number of photons in the
resonator.

To describe the transmission properties of the cavity as a function of the qubit state,
we write a closed set of differential equations from \req{eq:Hamiltonian0} describing
the time evolution of the system operators in the Heisenberg picture. These are deduced
from input-output equations established in the case of a transmitting cavity as in
\cite{christoph}. We define the external fields $b_{in}$ (injected microwave field),
$b_r$ (reflected field), and $b_t$ (transmitted field) that lead to the usual
input-output equations:
 $b_r = b_{in} + i \sqrt \kappa a$ and $b_t = i \sqrt \kappa a$ where
$\kappa$ is the resonator coupling to external transmission line modes.  As we consider
an overcoupled cavity, we neglect the internal losses of the resonator and thus $\kappa$
entirely defines the resonator linewidth. The qubit energy relaxation and dephasing
times are assumed to be very long compared to the resonator relaxation time ($\kappa
T_1\gg 1$ and $\kappa T_2\gg 1$). The Heisenberg equations are written in the frame
rotating at the frequency $\omega$ of the probe, yielding

\begin{equation}\begin{split}\label{eq:HeisenSystem}
\dot{\sigma_z^a } & = -2 g_a( \sigma_+^a  a+ \sigma_-^a a \dg) \, ,\\ %
\dot{\sigma_-^a} & = - i (\omega_r - \omega +  \delta_j ) \sigma_-^a + g_a \sigma_z^a a \, ,\\ %
\dot{\sigma_-^{qb}} & =  - i (\omega_{qb} -  g_{zz} \sigma_z^a ) \sigma_-^{qb}  \, ,\\ %
\dot{a} & = -i (\omega_r - \omega) a - \kappa a + g_a \sigma_-^a + i \sqrt \kappa b_{in}  \, ,\\ %
\end{split}\end{equation}
where $\delta_j = - g_{zz} (1 +\sigma_z^{qb})$ is the qubit state dependent shift, and
the index $j$ defines the qubit state ($j=g$ or $e$). As expected from a QND
measurement, the evolution preserves $\avg{\sigma_z^{qb}}$. We are interested in the
transmission properties of this system in the steady state regime established after a
time much larger than $1/\kappa$. We adopt the semiclassical approach where the quantum
correlations between atomic and field operators are neglected \cite{christoph}. From now
on we identify the operators with their average complex values, which could be measured
in a homodyne experiment. The ratio $t(\omega) = \avg{b_t} / \avg{b_{in} } $ can be
written, in the steady state regime
\begin{equation}\label{eq:transmission}
t_j(\omega)\! = t_0(\omega) \curlybraket{ 1 - \frac{1}{1 + \frac{p}{p_s} } \squarebraket{1 - \frac{2i (\omega_r + \delta_j - \omega)}{ \Gamma t_0(\omega)}}^{-1} }
\end{equation}
where $\Gamma= 2 g_a^2 / \kappa$, and $t_0(\omega) = - [ 1 + i (\omega_r-\omega) /
\kappa ]^{-1} $  is the transmission of the empty resonator. We have introduced the
drive power in units of photons per second $p = \avg{b_{in}\dg b_{in} }$, and the
saturation power $p_s$ of the atom-cavity system reads

$\frac{p_s}{\Gamma} =  \frac{(\omega_r + \delta_j - \omega)^2}{\Gamma^2}  + \squarebraket{ \frac{ (\omega_r-\omega)}{\Gamma} \frac{(\omega_r + \delta_j - \omega)}{\kappa} -1/2 }^2$.

\begin{figure}[htb]
\begin{center}
\includegraphics[width=7cm,height=4.2cm]{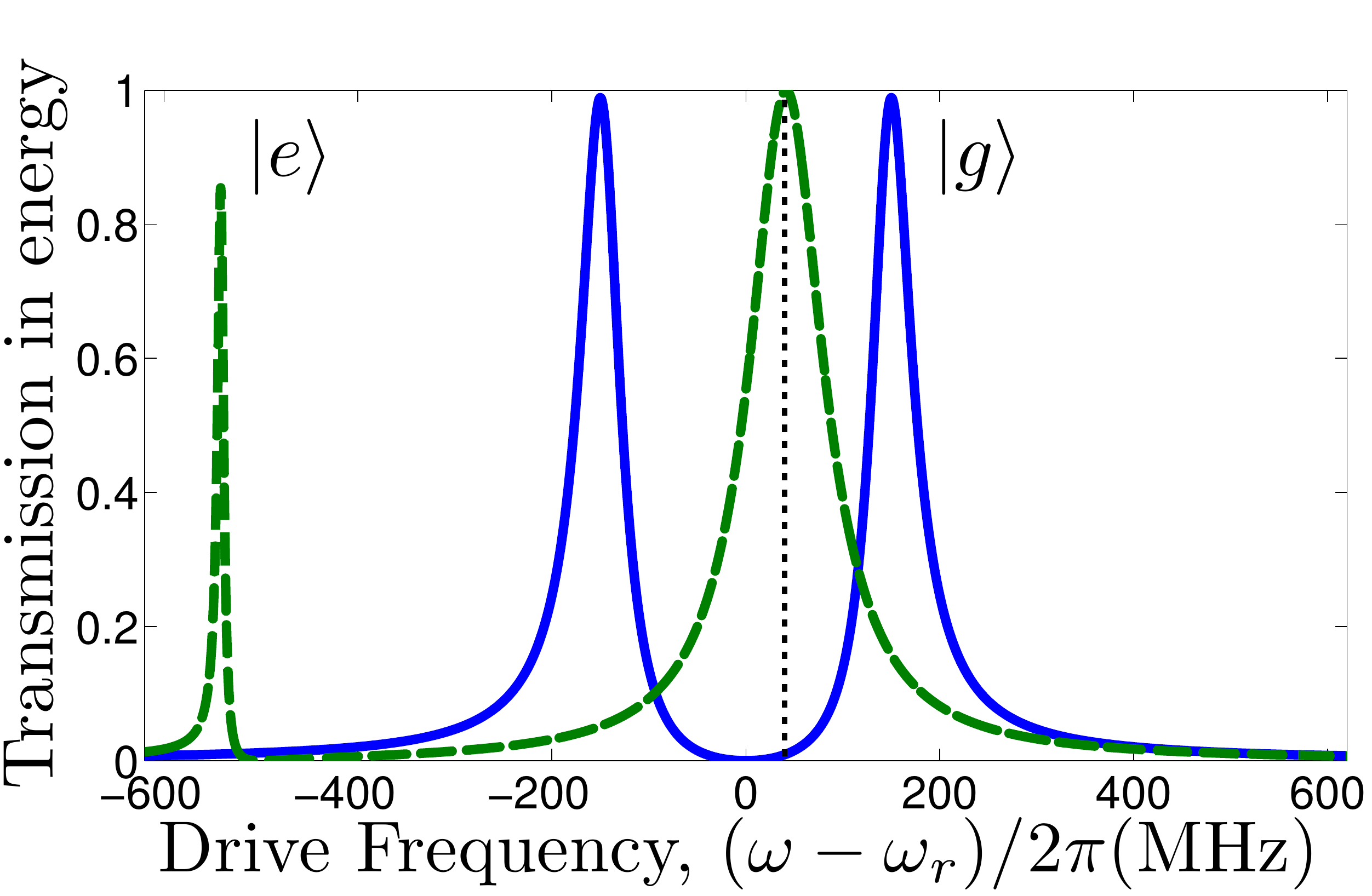}
\caption{Transmission coefficient at low pump power for a microwave pulse injected in a
 resonator containing a diamond-shape artificial atom. Blue solid curve: Qubit state
$\ket{g}$. Green dashed curve: Qubit state $\ket{e}$. Frequency is centered on the bare
cavity. We took $g_{zz}/2\pi  =250$ MHz , $g_a/2\pi=150$
MHz and a cavity linewidth $\kappa/2\pi =40$ MHz ($Q = 250$) and $p<<p_s$. The black
dotted line indicates the pump frequency $\omega$. } \label{fig:2}
\end{center}
\end{figure}

The essence of the protocol is pictured in Fig.\ref{fig:2}, in the linear regime when $p
\ll p_s$. In this regime the transmission is given by $t_j(\omega) = \left[
\frac{1}{t_0(\omega)} + \frac{i \Gamma}{ 2(\omega_r + \delta_j - \omega)} \right]^{-1}$.
If the qubit is in state $\ket{g}$, $\delta_g=0$ so that the ancilla qubit is resonant
with the cavity mode and the transmission consists on two peaks located at $\pm g_a$
with respect to the frequency of the resonator. If the qubit is in state  $\ket{e}$, $|
\delta_e |=2g_{zz}$, inducing a dispersive coupling between the resonator and the
ancilla provided that $g_{zz} > g_a$. The transmission essentially consists in a single
peak slightly shifted by $\delta_L= g_{zz} ( \sqrt{1 + g_a^2/ g_{zz}^2 } -1)$ with
respect to $\omega_r$ (see Fig. \ref{fig:2}). Thus a change in the state of the qubit
can now translate into a switch from dispersive to resonant coupling between the
resonator and the ancilla. This is evidenced by a visible displacement in the
transmission peaks by a quantity $g_{a}$, which can be as high as $150MHz$, about two
orders of magnitude higher than the usual dispersive ac-Stark
shift\cite{Vijay_PRL2011,Riste_PRL2012}. This strong effect allows an increase in the
linewidth of the resonator while keeping a high fidelity read-out. Working with a low Q
cavity has important advantages. First, it drastically increases the total bandwidth of
the circuit, and consequently the read-out speed. Moreover, for the same probe power,
the average intracavity photon number is lower, preserving the lifetime and coherence
time of the qubit\cite{Boissonneault_PRA2009}. The read-out is performed by the
injection of a short microwave pulse  of power $p$ at the frequency
$(\omega_r+\delta_L)/2\pi$. Thus, the transmitted power depends on the state of the
qubit, giving rise to two conditional output signals $p_{t|j} =\avg{b_{t}\dg b_{t} } =
|t_j|^2 p$. When $p$ largely overcomes $p_s$, one recovers the transmission pattern
$t_0(\omega)$ of the empty cavity, a signature of saturation \cite{christoph} which
limits the information on the qubit state.

We now introduce the model to optimize the measurement scheme. The performance of the
read-out is usually quantified by two figures of merit, namely their fidelity ${\cal F}$
and speed. Speed is high when the system can be measured frequently, the delay between
two measurements being inferiorly bounded by their typical correlation time $\tau_c$.
$\tau_c$ is related to the inertia of the circuit, since the resonator imposes
$\tau_c>\kappa^{-1}$. Fidelity and correlation time depend on two independent parameters
to optimize. First, the resonator linewidth should be narrow enough to give a large
contrast between the two transmission patterns ($\kappa < g_a,g_{zz}$), while large
enough to allow a large transmitted signal $p_{t|j}=\bar{n}_j \kappa$ and therefore high
speed qubit read-out for a given photon number $\bar{n}_j$ inside the resonator. In the
same way, the driving power $p$ should be sufficiently low to avoid the saturation of
the ancilla $p<<p_s$, but high enough to have a large $p_{t|j}$.

In a typical circuit QED experiment, microwave photons are amplified before being sent
through a homodyne detection scheme and digitalized within a short time interval $\tau$,
which is usually equal to $\tau_c$. For our purpose, we shall consider the field at the
entrance of the amplifying chain. The chain is modeled by a perfect amplifier
\cite{Bergeral_Nature2010} radiating at the input of the circuit a white thermal field
of effective temperature $T_N$. This noise temperature ranges from a few hundreds of mK
for the recent generation of quantum limited devices\cite{Roch_PRL2012,Johnson_PRL2012},
to $4-10$K for commercial devices. The total noise power ${\cal N}=(k_B T_N/\hbar
\omega)B$, in units of photons per second, is given by Johnson-Nyquist
noise\cite{Johnson&Nyquist_PR1928},  where $B$ is the bandwidth of the amplifier,
imposing an additional lower band to the correlation time $\tau_c>B^{-1}$. Consequently,
high speed measurements are obtained at the price of increased bandwidth and noise power
${\cal N}$.

Estimation of the  read-out fidelity is based on the photon number distributions
$\textbf{P}(n|j)$ conditioned to the qubit $j$, which we computed using the
Glauber-Sudarshan P-representation \cite{Glauber1993}. In our case, this simply
corresponds to the P-representation of a thermal field of temperature $T_N$ displaced by
a coherent field of amplitude $\sqrt{p_{t|j} }$. Thus we can readily calculate the
generating function \cite{MandelBook} for the photon statistics, from which we extract
the coefficients
\begin{equation}\begin{split}\label{eq:GeneratingFunction}
\textbf{P}(n|j)= \frac{\mathcal{N}^n \tau^n}{(1+ \mathcal{N}\tau)^{n+1}}
\exp\roundbraket{ \frac{ - \tau \, p_{t|j} }{1+ \mathcal{N}\tau} } L_n \roundbraket{
\frac{- p_{t|j} \; / \mathcal{N} }{1+\mathcal{N}\tau } } ,
\end{split}\end{equation}
where $L_n$ is the $n$-th order Laguerre polynomial.

\begin{figure}[htb]
\begin{center}
\includegraphics[width=9cm]{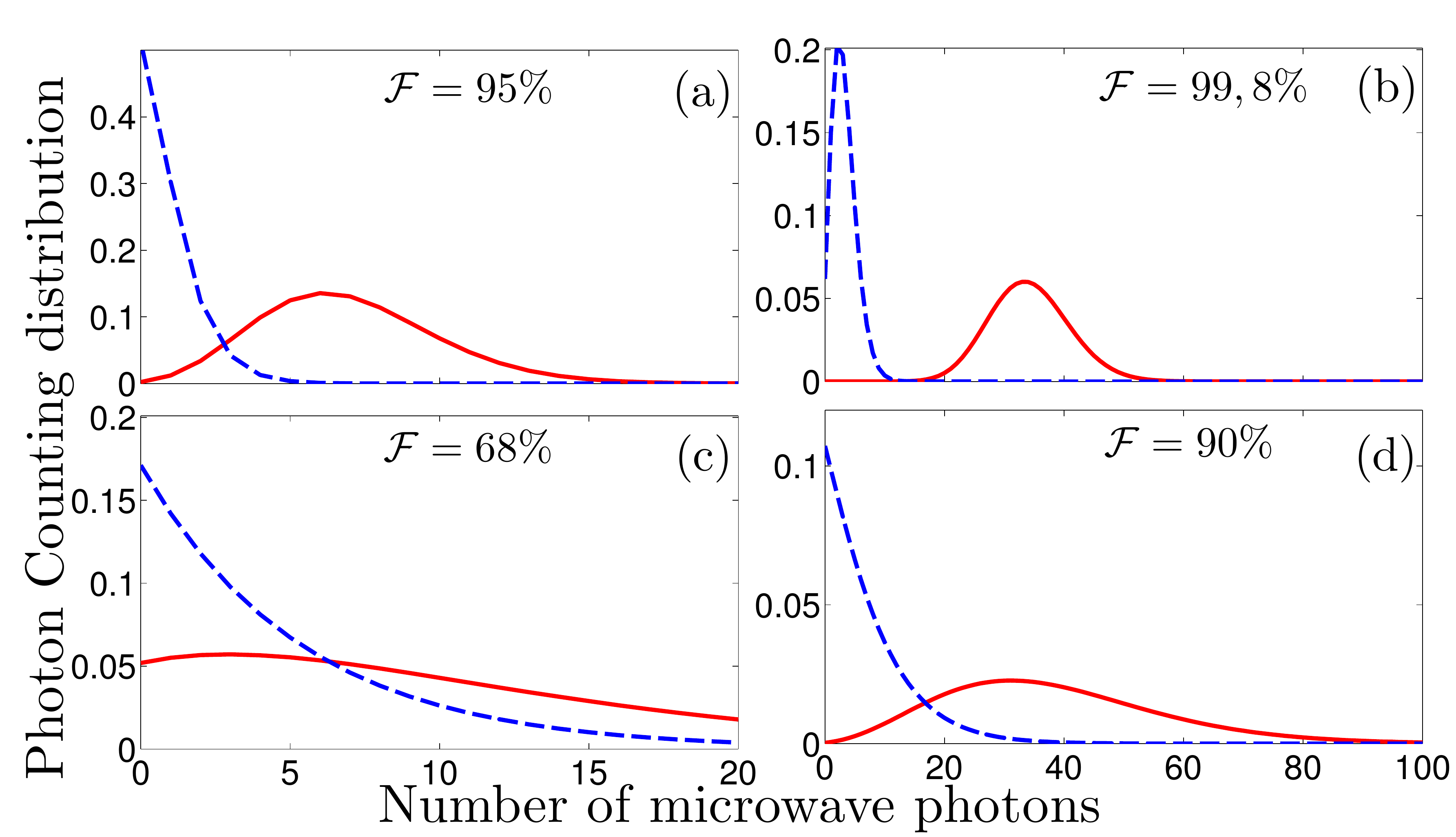}
\caption{Photon distribution at the entrance of the amplifier. Histograms $\textbf{P}(n|
e)$ (red solid) and $\textbf{P}(n| g)$ (blue dashed) with $\kappa =40MHz$ and $p=1
photon/ns$. (a) $T_N =  140$mK , $\tau= 10$ns, $B=50$MHz ; (b) $T_N =  140$mK , $\tau=
50$ns, $B=10$MHz ; (c)$T_N =  4$K , $\tau= 10$ns, $B=50$MHz ; (d) $T_N =  4$K , $\tau=
50$ns, $B=10$MHz} \label{fig:3}
\end{center}
\end{figure}

The histograms plotted in Fig. \ref{fig:3} clearly show how the amplification noise has
a large effect on the statistics of the counts associated with each of the qubit states.
As expected, the noise power increases with noise temperature (a vs. c and b vs. d),
degrading the fidelity. By increasing the integration time, one can regain fidelity (a
vs. b and c vs. d). As a matter of fact, it increases the signal, but it also allows one
to operate with a lower bandwidth, reducing the noise power. We expect this protocol to
yield a fidelity as high as $90\%$ with a commercial amplifier, within a typical time of
$\tau=50$ns. An integration as short as $\tau=60$ns should be enough to reach $99.9 \%$
with a state of the art amplifier with $T_N=140$mK.

\begin{figure}[htb]
\begin{center}
\includegraphics[width=8cm]{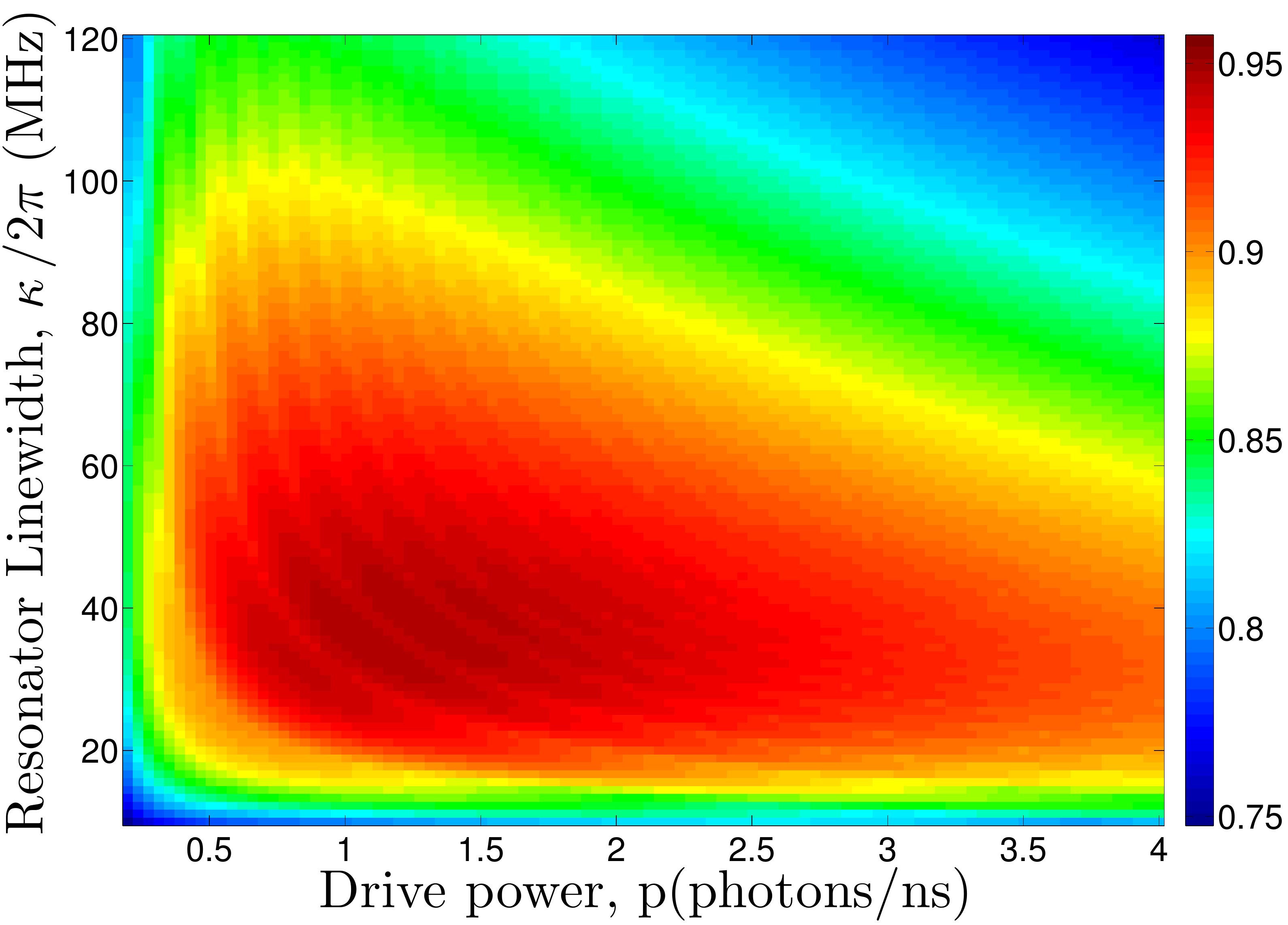}
\caption{Single measurement fidelity for a state of the art amplifier ($T_N= 140$mK) and an
acquisition time $\tau=10$ ns versus drive power $p$ in units of photons/$10$ns and
resonator linewidth $\kappa$. Optimal value of ${\cal F}=95\%$ is reached for broad range
around $\kappa=40$MHz and $p=1 photon/ns$.} \label{fig:4}
\end{center}
\end{figure}

Fig. \ref{fig:4} shows optimization of the fidelity as a function of the resonator
linewidth and probe power. The digitization time $\tau = 10$ns has been chosen,
compatible with a bandwidth $B=50$MHz. A fidelity ${\cal F}=95\%$ can be reached with a
resonator linewidth $\kappa = 40$ MHz and very small pumping power, corresponding to
$\bar{n} = 1.8$ photons. This fidelity corresponds to up to date results obtained in the
dispersive measurement scheme with the same amplifier \cite{Johnson_PRL2012,
Vijay_PRL2011}, but allows a much faster acquisition time. Indeed, in dispersive based
read-out schemes the dynamics is slow because of the inertia imposed by the resonator of
linewidth $\kappa\sim 5$MHz, inducing a typical correlation time of $\tau_c=100$ns. With
our scheme, using a low temperature amplifier allows a drastic increase in the bandwidth
and read-out speed. This enables a projective measurement of the qubit to be performed
on a timescale much shorter than the recently measured relaxation time, $T_1=50\mu s$
\cite{Paik_PRL2011}. This scheme opens the path to the observation of quantum jumps in
circuit QED with a very high temporal resolution, comparable to the performances
achieved in recent experiments performed with Rydberg atoms \cite{Gleyzes2007}, where
the system is typically measured $10^3$ times before undergoing a quantum jump.

In conclusion we propose a new read-out scheme based on a superconducting diamond-shape
artificial atom which contains a logical qubit strongly coupled to an ancilla qubit by a
cross-Kerr term. We predict fast high-fidelity  QND read-out of the transmon qubit with
a commercial amplifier. Using a quantum limited amplifier, $60$ ns read-out time and
$99.9\%$ fidelity are predicted. This original method overcomes the current read-out
limitation of the superconducting qubits dispersively coupled to a resonator. In
addition Purcell effect between the logical qubit and cavity is absent. As a side effect
the intra-cavity population is minimal, $\bar{n}=1.8$ for the optimal parameters,
minimizing any adverse effect on the qubit coherence properties. This opens the
possibility of monitoring quantum jumps of the qubit with very high temporal resolution
\cite{Gleyzes2007}, of generating non-classical states \cite{Zhou2012} or implementating
quantum error correction codes \cite{Chuang} using closed feedback loops.

The authors gratefully thank N. Roch, P. Bertet, P. Milman , A. K. Feofanov and I. M.
Pop for useful discussions. This work was supported by the European SOLID and the
ANR-NFSC QuExSuperC projects, Nanosciences Foundation of Grenoble and CAPES.

\end{document}